\documentclass[twocolumn]{aastex6}

\newcommand{\Msun}{\ifmmode M_{\odot} \else $M_{\odot}$\fi}
\newcommand{\Msunyr}{\ifmmode M_{\odot}\, {\rm yr}^{-1} \else $M_{\odot}\, {\rm yr}^{-1}$}

\begin{document}

\title{A comparison of two methods for estimating black hole spin in active
galactic nuclei}

\author{Daniel M. Capellupo\altaffilmark{1}}
\affil{Department of Physics, McGill University, Montreal, QC, H3A 2T8, Canada \\
McGill Space Institute, McGill University, Montreal, QC, H3A 2A7, Canada}

\author{Gaylor Wafflard-Fernandez}
\affil{Department of Physics, Universit\'{e} Paris-Sud, Orsay, France}

\author{Daryl Haggard}
\affil{Department of Physics, McGill University, Montreal, QC, H3A 2T8, Canada \\
McGill Space Institute, McGill University, Montreal, QC, H3A 2A7, Canada}

\altaffiltext{1}{danielc@physics.mcgill.ca}

\begin{abstract}

Angular momentum, or spin, is a fundamental property of black holes (BHs), yet
it is much more difficult to estimate than mass or accretion rate (for actively
accreting systems). In recent years, high-quality X-ray observations have
allowed for detailed measurements of the Fe K$\alpha$ emission line, where
relativistic line broadening allows constraints on the spin parameter (the
X-ray reflection method). Another technique uses accretion disk models to fit
the AGN continuum emission (the continuum-fitting, or CF, method). Although
each technique has model-dependent uncertainties, these are the best empirical
tools currently available and should be vetted in systems where both techniques
can be applied. A detailed comparison of the two methods is also useful because
neither method can be applied to all AGN. The X-ray reflection technique
targets mostly local (z $\lesssim$ 0.1) systems, while the CF method can be
applied at higher redshift, up to and beyond the peak of AGN activity and
growth. Here, we apply the CF method to two AGN with X-ray reflection
measurements. For both the high-mass AGN, H1821+643, and the Seyfert 1,
NGC 3783, we find a range in spin parameter consistent with the X-ray
reflection measurements. However, the near-maximal spin favored by the
reflection method for NGC 3783 is more probable if we add a disk wind to the
model. Refinement of these techniques, together with improved X-ray
measurements and tighter BH mass constraints, will permit this comparison in a
larger sample of AGN and increase our confidence in these spin estimation
techniques.

\end{abstract}


\section{Introduction} \label{sec:intro}

Actively accreting black holes have three fundamental properties -- mass
($M_{BH}$), accretion rate ($\dot{M}$), and angular momentum. Measuring
$M_{BH}$ for active galactic nuclei at all redshifts has become possible due to
reverberation mapping of low-redshift AGN and the extrapolation of those
results to high redshifts, via relations between $M_{BH}$ and the widths of
broad emission lines and the AGN continuum luminosity. Accretion rate estimates
have also been achieved for many AGN, usually via the Eddington ratio,
$L/L_{Edd}$.

The angular momentum, or spin ($a_*$), of active BHs is more elusive, as it
requires probing the region near the inner edge of the accretion disk (AD). Yet
measurements of spin and spin evolution would provide valuable clues to the
accretion history of active BHs and perhaps the evolution of the AGN and host
galaxies themselves.

At present, there are two primary methods for constraining the spin parameters
of actively accreting BHs: (1) measuring the Fe K$\alpha$ emission line and/or
a soft X-ray excess that some attribute to relativistic reflection
\citep[e.g.][]{Brenneman13,Reynolds14b}, and (2) fitting the AGN continuum
emission (CF) \citep[e.g.][]{Done13,Capellupo16}. There are significant
advantages and drawbacks to each method.

The Fe K$\alpha$ method is based on relativistic X-ray reflection. It does not
require prior knowledge of $M_{BH}$, the distance to the source, or the
inclination of the disk, whereas these are all necessary ingredients for the CF
method. The main drawback, however, is that a very high-quality X-ray spectrum
is required to properly model the continuum emission and the Fe-K$\alpha$
emission line, severely limiting the number of sources for which current
technology allows a spin measurement. As a result, most AGN with reflection
measurements are at a redshift less than 0.1. Furthermore, the Fe K$\alpha$
emission line is present in just $\sim$40\% of bright, nearby type I AGN
\citep{deLaCallePerez10}, so some spin estimates are based on modeling just a
soft X-ray excess \citep[e.g.][hereafter, R14]{Reynolds14a}.

The CF method, on the other hand, can be applied to any AGN where the continuum
emission can be measured. This vastly increases the number of AGN for which a
spin measurement can be made and has already been applied out to a redshift of
$\sim$1.5 \citep{Capellupo15,Capellupo16}. The primary drawback is that wide
wavelength coverage, sometimes exceeding the capabilities of a single
observatory, is required to properly measure the shape of the SED. Furthermore,
this method cannot be applied effectively if the peak of the AD spectrum occurs
in a wavelength regime inaccessible to current observatories, e.g., the
extreme UV (where many AGN spectra do indeed peak). This method generally
assumes a thin AD model, based on \citet{Shakura73}.

Recent work has directly cast doubt on the X-ray reflection method.
\citet{Boissay16} find that the soft X-ray excess that some attribute to
relativistic reflection is more likely due to warm Comptonization. Similarly,
\citet{Yaqoob16} is able to fit the Fe K$\alpha$ emission line for one of the
AGN with an X-ray reflection spin measurement without invoking relativistic
reflection. For the CF method, while the standard thin AD model has been
successful in fitting the UV-optical SEDs of many AGN
\citep[see e.g.][]{Capellupo15,Capellupo16}, other work has found that the AGN
SED can be fit with the combination of a thermal disk component and a warm
Comptonization component \citep[][]{Mehdipour11}, indicating the possibility of
greater complexity in the continuum emission.

With these two methods now available and actively in use for the estimation
of $a_*$ in AGN, it is time to investigate whether these two methods give
consistent results when applied to the same AGN. This is especially important
given the uncertainties in both techniques and because neither method can probe
the full AGN population.

In this work, we compare the X-ray reflection and CF methods for two nearby AGN
-- H1821+643 and NGC 3783. Ours is among the first attempts to make this
comparison \citep[see also,][]{Done16}. Both targets have a published spin
estimate from the reflection method. We perform the CF analysis and compare the
results in detail. In \S2, we describe how we selected sources for this study
and our search for appropriate archival data. In \S3, we describe the models
and CF procedure (based on \citealt{Capellupo15,Capellupo16}). \S4 and \S5
describe our application of the CF method to the two AGN, and we conclude in
\S6 with a discussion of our results and how the reflection and CF method
compare for these two case studies. We assume a $\Lambda$CDM model with
$\Omega_{\Lambda}=0.7$, $\Omega_{m}=0.3$, and
$H_{0}=70\, {\rm km\, s^{-1}} \, {\rm Mpc}^{-1}$.

\section{Sample Selection and Data Sources} \label{sec:data}

According to \citet{Vasudevan16}, there are currently 25 AGN with spin
estimates from the X-ray reflection method. We use this list as a starting
point to search for archival data to which the CF method can be applied.

The CF method is most effective when the ``turnover'' in the AD spectrum is
probed. This turnover occurs at shorter wavelengths for smaller black hole
masses. We therefore look first for existing high-quality UV spectroscopic
observations of these AGN. Via the MAST web portal\footnote{\url{https://archive.stsci.edu/}},
we identify four AGN with high-level data products for Hubble Space Telescope
(HST) Faint Object Spectrograph (FOS) observations \citep{Evans04}: Fairall 9,
NGC 3783, NGC 4151, and H1821+643.

While the FOS spectrum is sufficient for applying the CF method for H1821+643,
data at even shorter wavelengths is required for the lower--$M_{BH}$ Seyfert
galaxies. We seek quasi-simultaneous data, and, for NGC 3783, we identify
observations from ROSAT -- taken on 1992 July 23, just four days prior to the
FOS observation on 1992 July 27 -- that probe the appropriate wavelengths
\citep{Alloin95}. Hence we proceed with two objects, H1821+643 and NGC 3783,
for our detailed spin comparison.

The FOS spectra for H1821+643 and NGC 3783 are focused on the nucleus of the
galaxy. For H1821+643, we verify that the FOS spectrum (shown in
Fig.~\ref{fig:Hspec}) is dominated by AGN emission based on the broad-band star
formation SED fit in \citet{Farrah02}. Similarly for NGC 3783, the spectrum is
at short enough wavelengths that the host galaxy contribution should be
negligible \citep{Reichert94,Alloin95}. Therefore, we do not correct the FOS
spectra for stellar emission.

The only correction we make to the HST data is to divide out the Galactic
extinction, using the \citet{Cardelli89} extinction law and the
\citet{Schlafly11} recalibration of the \citet{Schlegel98} maps.

The ROSAT data have been analyzed \citep[][hereafter, T93]{Turner93}, and we
make no further corrections in this work.

\section{Accretion Disk Models and Bayesian Routine} \label{sec:model}

To apply the CF method, a model is required that can make specific predictions
for the emitted radiation at each wavelength. Standard thin AD theory
\citep{Shakura73} has been used for several decades to describe AGN continuum
emission. Newer models use this framework, but incorporate general relativistic
corrections, comptonization in the disk atmosphere, and even disk winds
\citep[e.g.][]{Hubeny01,Davis11,Slone12}. Here we adopt the numerical code
described in \citet{Slone12}, assuming a viscosity parameter ($\alpha$) of 0.1.

The shape and luminosity of the thin AD spectrum is mainly set by $M_{BH}$,
$\dot{M}$, $a_*$, and the inclination of the disk to our line-of-sight.
If we want to constrain $a_*$, prior knowledge of the other parameters is
necessary as the observed SED is not enough to break the parameter degeneracy
of the models, where different combinations of these parameters can yield
similar SED shapes. Additionally, any intrinsic reddening in the AGN host
galaxy will affect the observed SED shape.

We therefore adopt a Bayesian approach that takes a large grid of models --
with varying values of $M_{BH}$, $\dot{M}$, $a_*$, inclination, and reddening
-- and maximizes the probability that any given model is a good representation
of the data, while penalizing those models that are not consistent with the
priors, which we establish for $M_{BH}$ and $\dot{M}$
\citep{Capellupo15,Capellupo16}. This routine calculates a $\chi^2$ value for
each model, using continuum windows along the observed SED.

For the prior on $M_{BH}$, the reverberation mapping technique has been used to
obtain $M_{BH}$ for nearby AGN \citep[e.g.,][]{Peterson04}, and these results
have been extended to other AGN, using the width of the broad emission lines
and the continuum luminosity
\citep[the `single-epoch method'; e.g.,][]{Bentz09,Mejia16}. A prior on
$\dot{M}$ can be estimated using $M_{BH}$ and a measurement of the continuum
luminosity at longer (i.e., optical or near-infrared) wavelengths, assuming
the canonical power law, $L_{\nu} \propto \nu^{1/3}$
\citep{Collin02,Davis11,Netzer14}.

For the disk inclination, the only constraint we have is that our sample
contains type-1 AGN, so we can consider only inclinations where
cos $\theta > 0.5$. For intrinsic reddening, to limit the number of free
parameters, we use a simple power-law curve, where
$A(\lambda)=A_{o}\lambda^{-1}$ mag. We consider values of $A_V$ ranging from
0.0 to 0.50 mag.

\section{H1821+643} \label{sec:h1821}

\begin{figure*}[]
\plotone{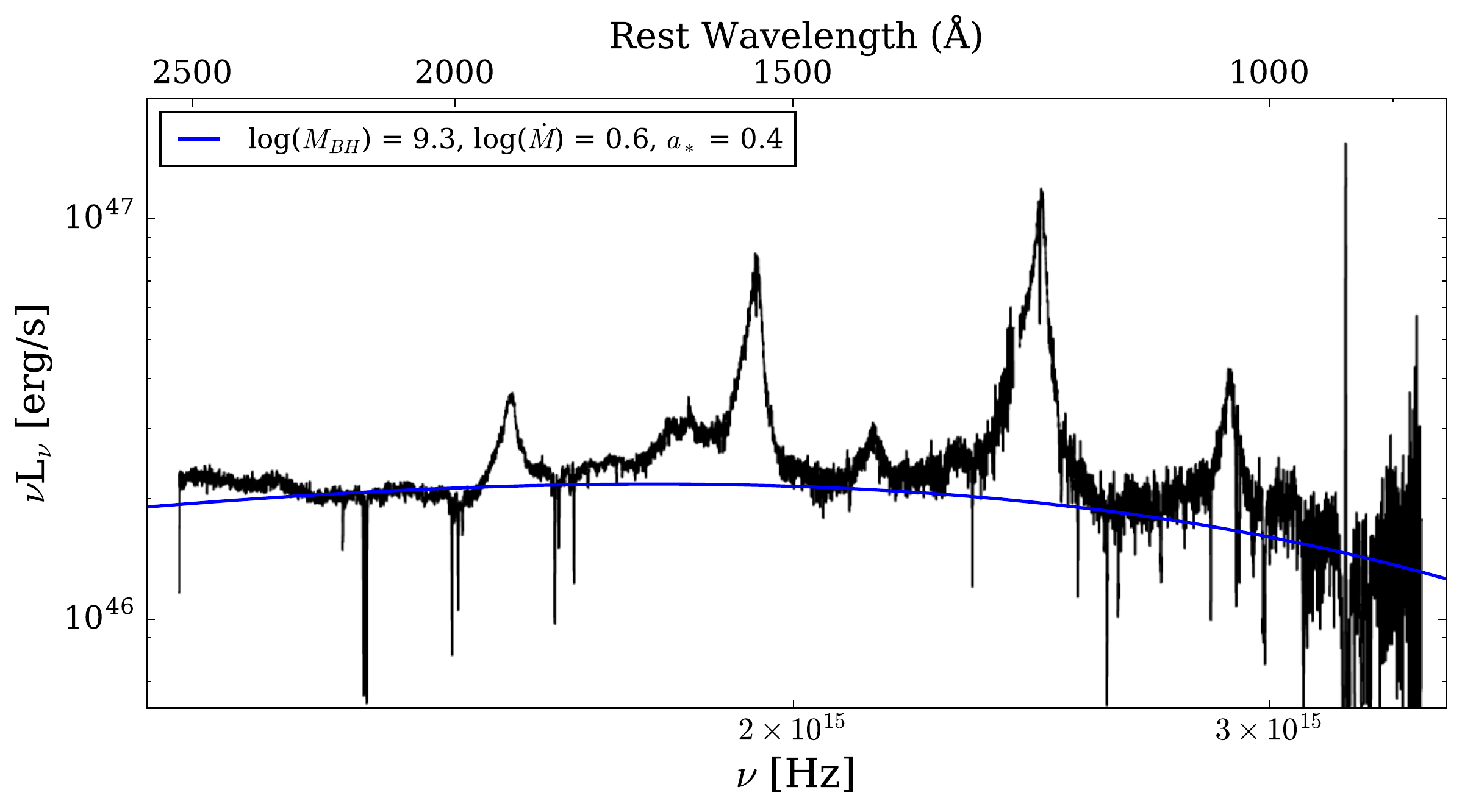}
\caption{The HST FOS spectrum (black curve) of H1821+643, with no intrinsic
  reddening correction. The best-fit CF model is overplotted.
  \label{fig:Hspec}}
\end{figure*}

H1821+643 is a brightest cluster galaxy (BCG) hosting a luminous AGN at
$z\sim0.297$. There are no direct reverberation mapping measurements for
H1821+643, but there have been several attempts to obtain $M_{BH}$ via other
methods. These estimates range from $\sim$1.2 to $6 \times 10^9 \Msun$
(\citealt{Decarli08,Dasyra11}; R14), and there are theoretical arguments that
the mass could be as high as $3 \times 10^{10}$ \Msun\ \citep{Walker14}. We
adopt the most recent `single-epoch' measurement using the H$\beta$ emission
line, $M_{BH} = 2.5 \times 10^9 \Msun$, from \citet{Decarli08}, and we use
their measurement of
log $\lambda L_{\lambda}(5100\mathrm{\AA}) =$ 46.1 ergs s$^{-1}$ for
calculating $\dot{M}$. We adopt errors of 0.3 and 0.2 dex, respectively, for
$M_{BH}$ and $\dot{M}$ \citep{Capellupo15}.

\begin{figure*}[]
\gridline{\fig{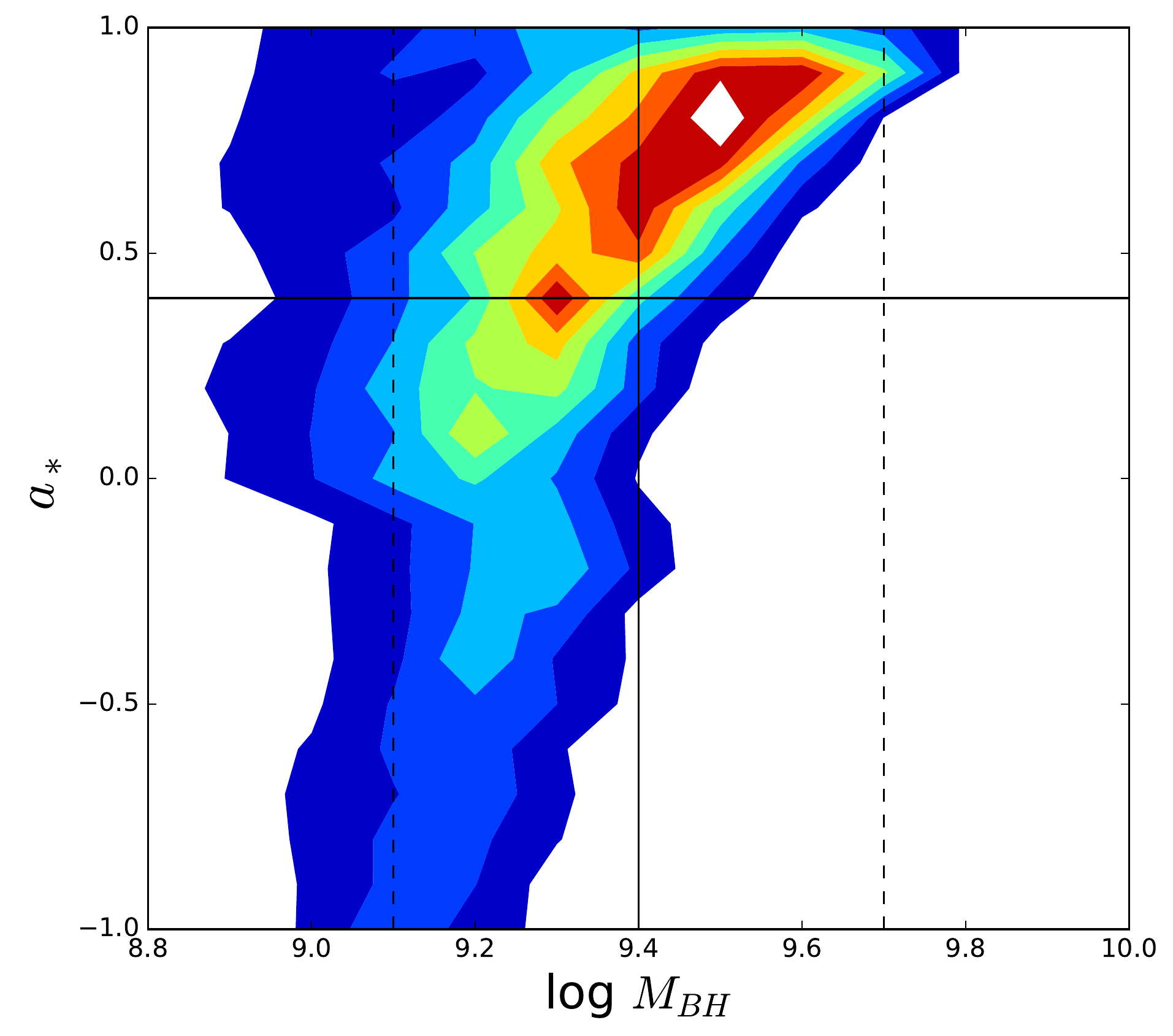}{0.40\textwidth}{}
		  \fig{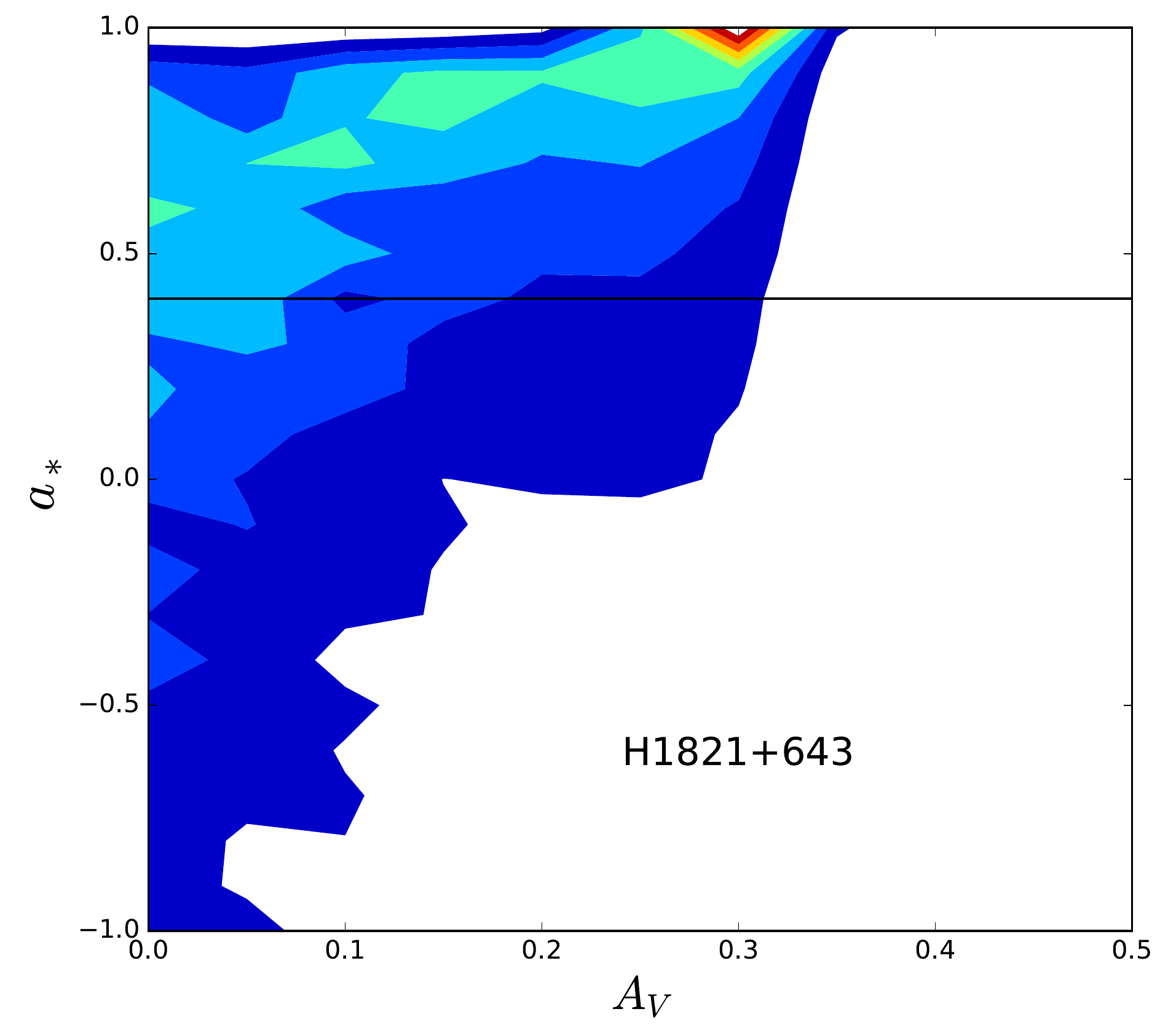}{0.40\textwidth}{}
		  \fig{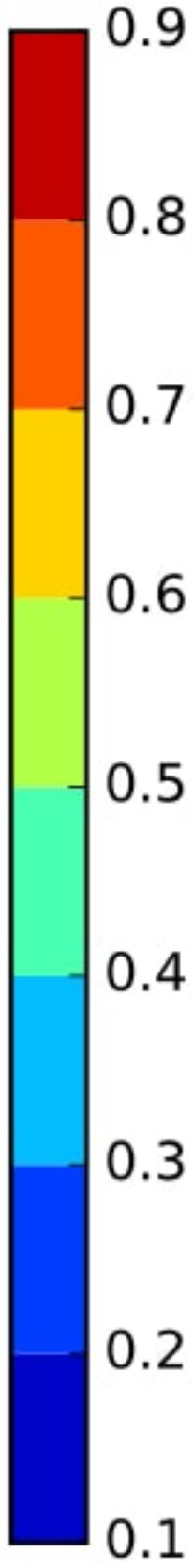}{0.045\textwidth}{}
          }
\caption{Posterior probability contour plots for H1821+643 for $a_*$ versus
  $M_{BH}$ and $a_*$ versus $A_V$. The vertical lines identify the observed
  $M_{BH}$ and the 0.3 dex error. The horizontal line identifies the lower
  limit on $a_*$ from R14.
  \label{fig:Hcont}}
\end{figure*}

\begin{deluxetable*}{c|cc|cccc|c}
\tablecaption{Model Parameters and Results \label{tab:table}}
\tablehead{
\colhead{Object} & \colhead{log $M_{BH}^{obs}$} & \colhead{log $\dot{M}^{obs}$} & \colhead{$L/L_{Edd}$} & \colhead{cos $\theta$} & \colhead{$A_V$} & \colhead{$a_{*}^{CF}$} & \colhead{$a_{*}^{ref}$} \\
\colhead{} & \colhead{(\Msun)} & \colhead{(\Msunyr)} &  &  & \colhead{(mag)} & \colhead{} & \colhead{}
}
\startdata
H1821+643                 & 9.4  & 0.48  & $0.14^{+1.8}_{-0.11}$     & $0.85^{+0.15}_{-0.09}$ & $0.12^{+0.15}_{-0.12}$  & $0.5^{+0.5}_{-0.4}$ & $\ge0.40$\tablenotemark{a}  \\
NGC 3783                  & 7.47 & -1.9  & $0.020^{+0.096}_{-0.014}$ & $0.89^{+0.11}_{-0.09}$ & $0.17^{+0.11}_{-0.09}$  & $0.2^{+0.7}_{-0.9}$ & $\ge0.88$\tablenotemark{b} \\
NGC 3783\tablenotemark{c} & 7.47 & -1.9  & $0.032^{+0.15}_{-0.018}$  & $0.90^{+0.10}_{-0.09}$ & $0.09^{+0.09}_{-0.06}$  & $0.5^{+0.5}_{-0.4}$ &            \\
\enddata
\tablenotetext{a}{R14}
\tablenotetext{b}{B11}
\tablenotetext{c}{CF with disc wind}
\end{deluxetable*}

The best-fit (i.e., the most probable) model is presented in
Fig.~\ref{fig:Hspec}, and the full results are shown as probability contours in
Fig.~\ref{fig:Hcont}. From Fig.~\ref{fig:Hcont}, it is clear that there is a
strong preference for a large, positive spin parameter.

In their analysis of the X-ray spectrum of H1821+643, R14 obtain both a
constraint on the spin parameter and a constraint on $L/L_{Edd}$ and the
inclination. Applying these constraints to our CF routine, we obtain a similar
probability distribution along the spin parameter axis as we did originally
without these constraints.

\section{NGC 3783} \label{sec:ngc3783}

\begin{figure*}[]
\plotone{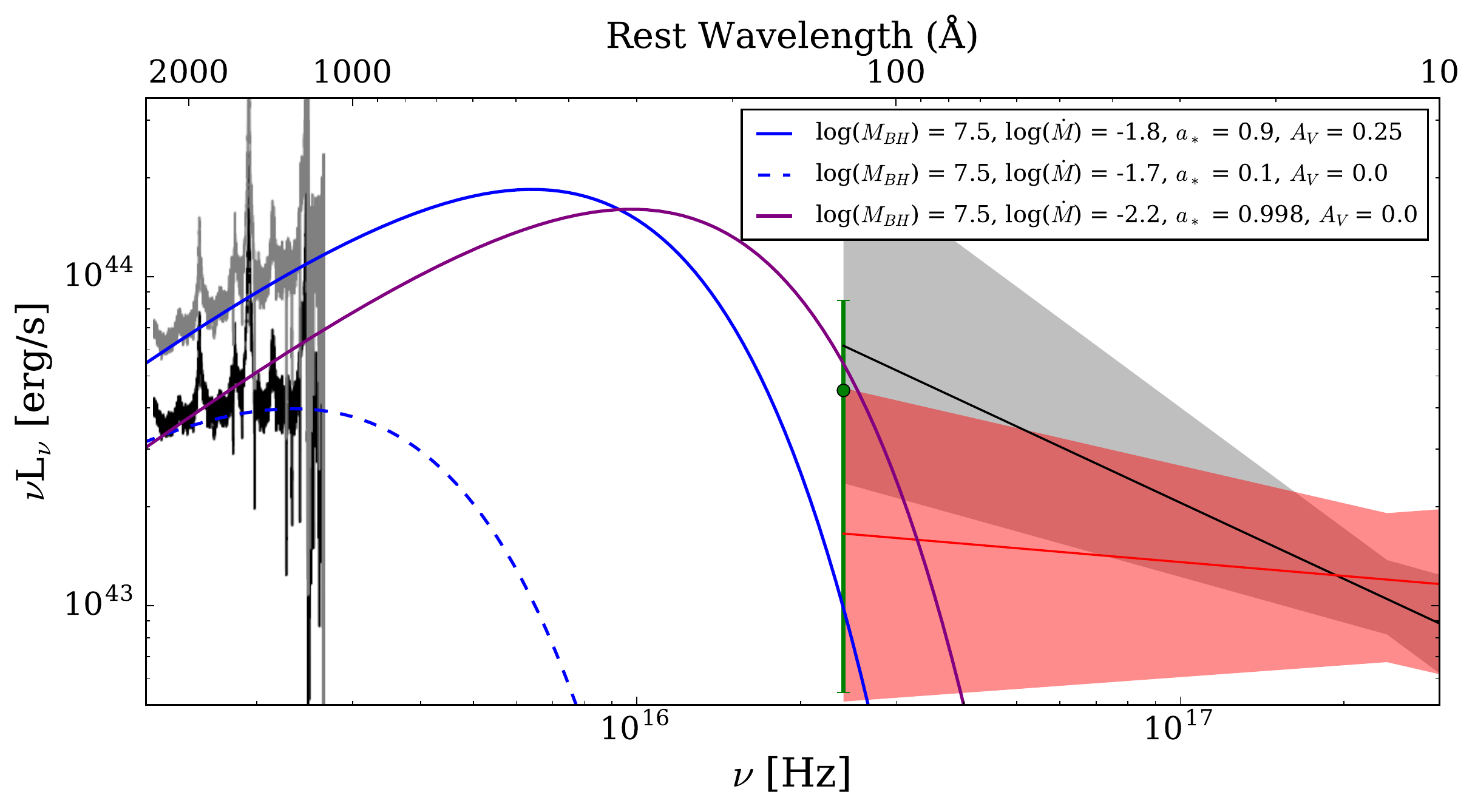}
\caption{The FOS spectrum (black curve) of NGC 3783, corrected for intrinsic
  reddening (gray curve), and the ROSAT (black) and EXOSAT (red) power-laws
  from T93, with shaded regions denoting the 1$\sigma$ error intervals. The
  solid and dashed blue curves are the best-fit models without and with a disk
  wind, respectively.
  \label{fig:Nspec}}
\end{figure*}

NGC 3783 is a well-studied Seyfert 1, SBa galaxy at $z\sim0.009$. The
reverberation mapping technique has been applied to NGC 3783, giving
$M_{BH}$ = $2.98 \pm 0.54 \times 10^7$ \Msun, with a corresponding continuum
luminosity,
log $\lambda L_{\lambda}(5100\mathrm{\AA}) = 43.26 \pm 0.04$ ergs $s^{-1}$,
which we use to estimate $\dot{M}$.

Because NGC 3783 is in a lower $M_{BH}$ regime than H1821+643, the peak of the
AD emission is in the extreme UV, a regime where we generally lack
observations. We can discriminate between different spin parameters only in the
soft X-ray, where models with the highest spin parameters peak for lower-mass
BHs.

NGC 3783 has a complex X-ray spectrum, with warm absorbers and a soft excess
that appears and disappears \citep{Netzer03}. We use ROSAT X-ray data (see
\S\ref{sec:data}), in addition to the FOS data, to apply the CF method to
NGC 3783. We use the 1992 July 23 ROSAT observation, in particular, because it
is nearly contemporaneous with the FOS observation, and it extends to slightly
lower energy (down to 0.1 keV) than more recent X-ray observations with Chandra
or XMM Newton. T93 fits the ROSAT data with several different power-laws based
on different absorption models, ranging from a simple power-law model with
$\Gamma$ = 2.22 to a warm absorber model with $\Gamma = 2.77^{+0.45}_{-0.31}$
(which is similar to the value found by \citealt{Schartel97} of
$\Gamma = 2.7 \pm 0.7$). T93 also present a model with $\Gamma \sim$ 4.7, which
is much higher than other values in the literature, so we do not include it in
our analysis.

\subsection{Applying the CF Method with an X-ray Upper-limit}

A difficulty with using the X-ray spectrum of an AGN for the CF method is that
there is a known power-law component at X-ray wavelengths of unknown origin, in
addition to possible emission from the AD. Hence, the X-ray data provides only
an upper limit on the AD emission.

We therefore first alter our CF method to search through our model parameter
space for the models with the highest spin parameter that give both a
satisfactory fit to the FOS spectrum ($\chi^2 \le 3$) and do not exceed the
X-ray flux at 0.1 keV from the power-law fits to the ROSAT data. To be
conservative in our upper limit, we adopt the warm absorber model power law
($\Gamma = 2.77^{+0.45}_{-0.31}$) from T93.

We find models spanning the full range in spin parameter, including maximum
spin, that can fit within the upper limit from the T93 warm absorber model
power-law for the ROSAT data, as long as $M_{BH}$ is at least as high as the
\citet{Peterson04} $M_{BH}$ estimate (see, for example, the purple curve in
Fig.~\ref{fig:Nspec}).

\subsection{Applying the CF Method with a Modified X-ray Flux}

\begin{figure*}
\gridline{\leftfig{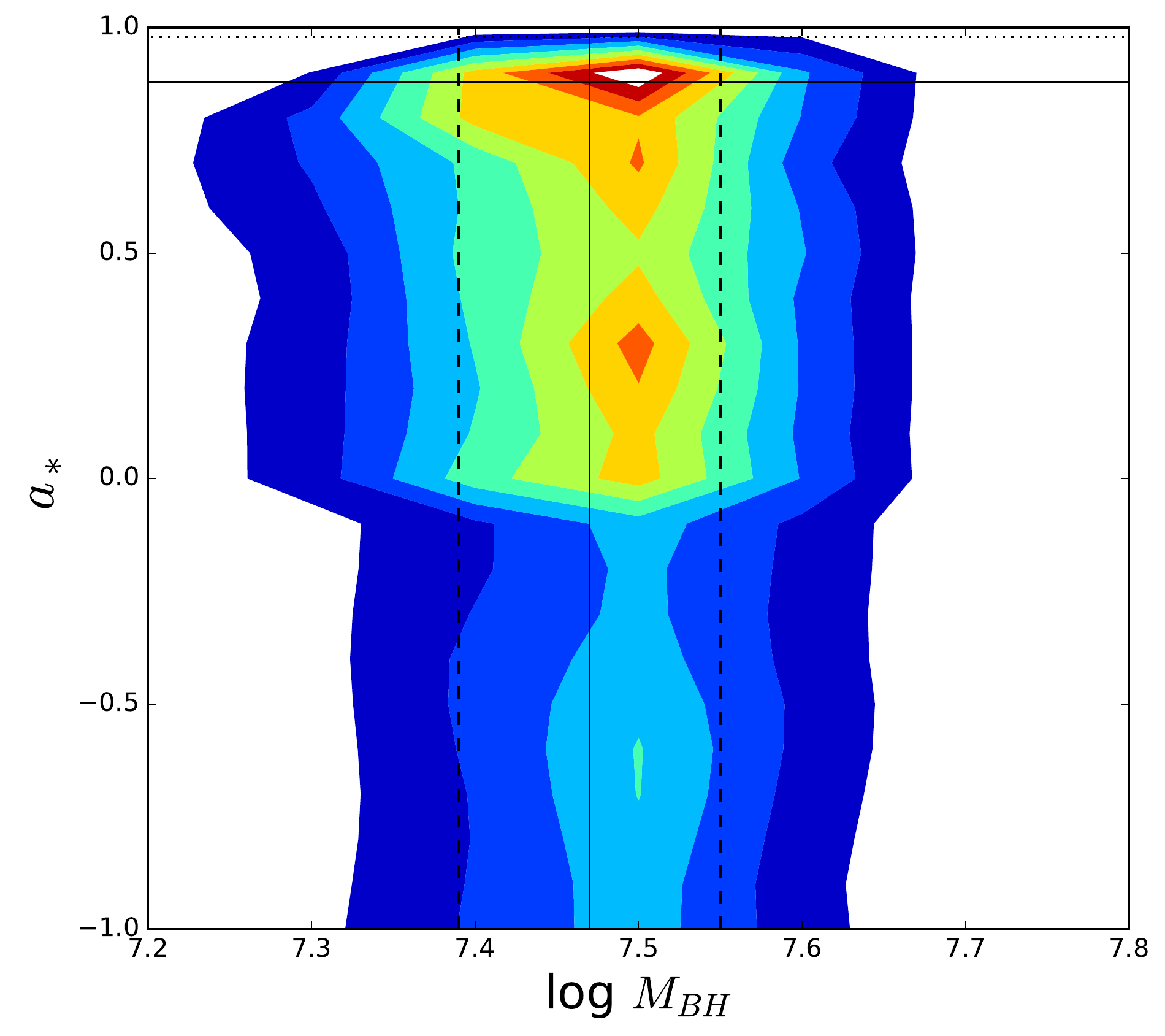}{0.45\textwidth}{}
		  \fig{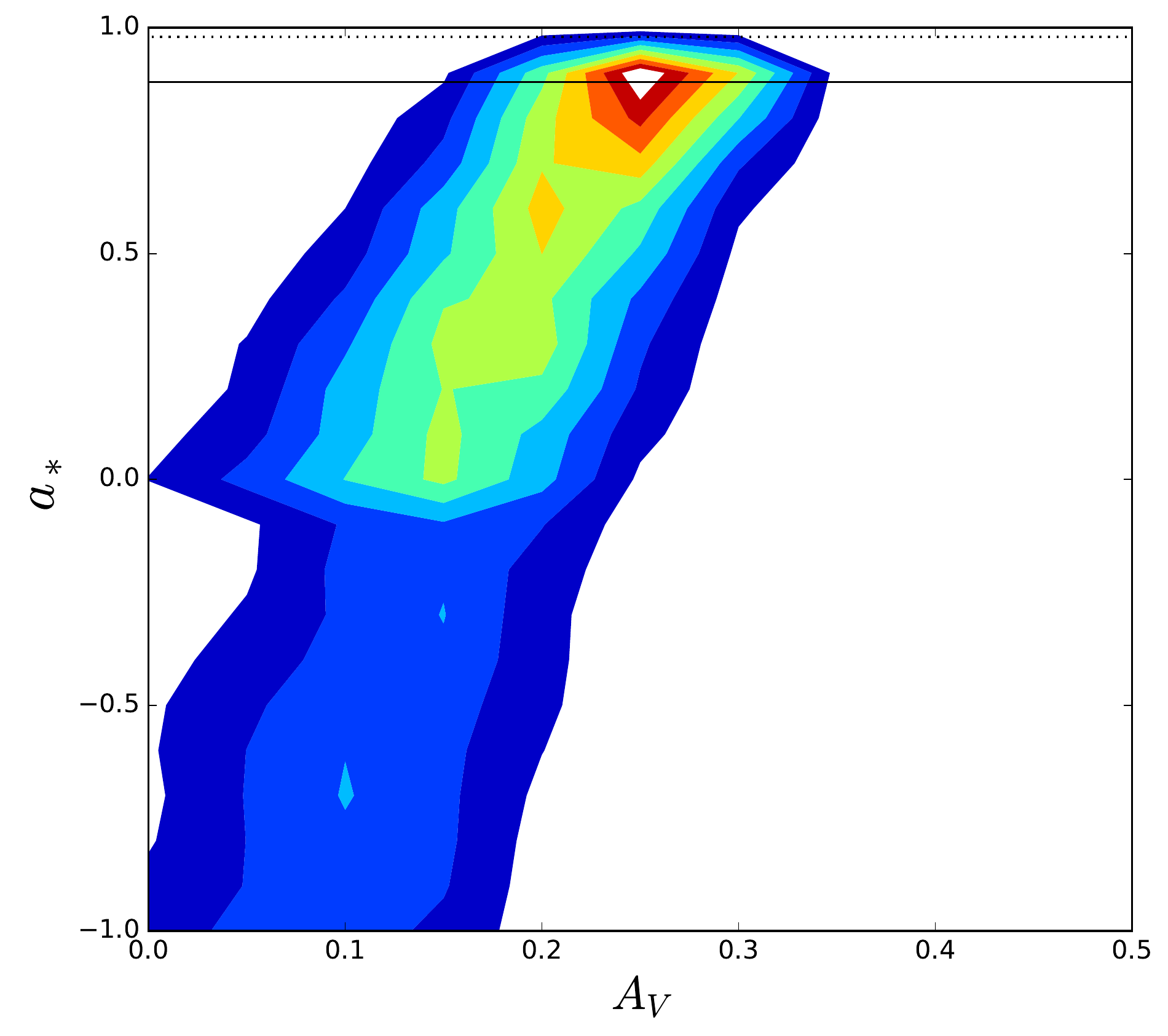}{0.45\textwidth}{}
		  \fig{fig2c.pdf}{0.05\textwidth}{}
          }
\gridline{\leftfig{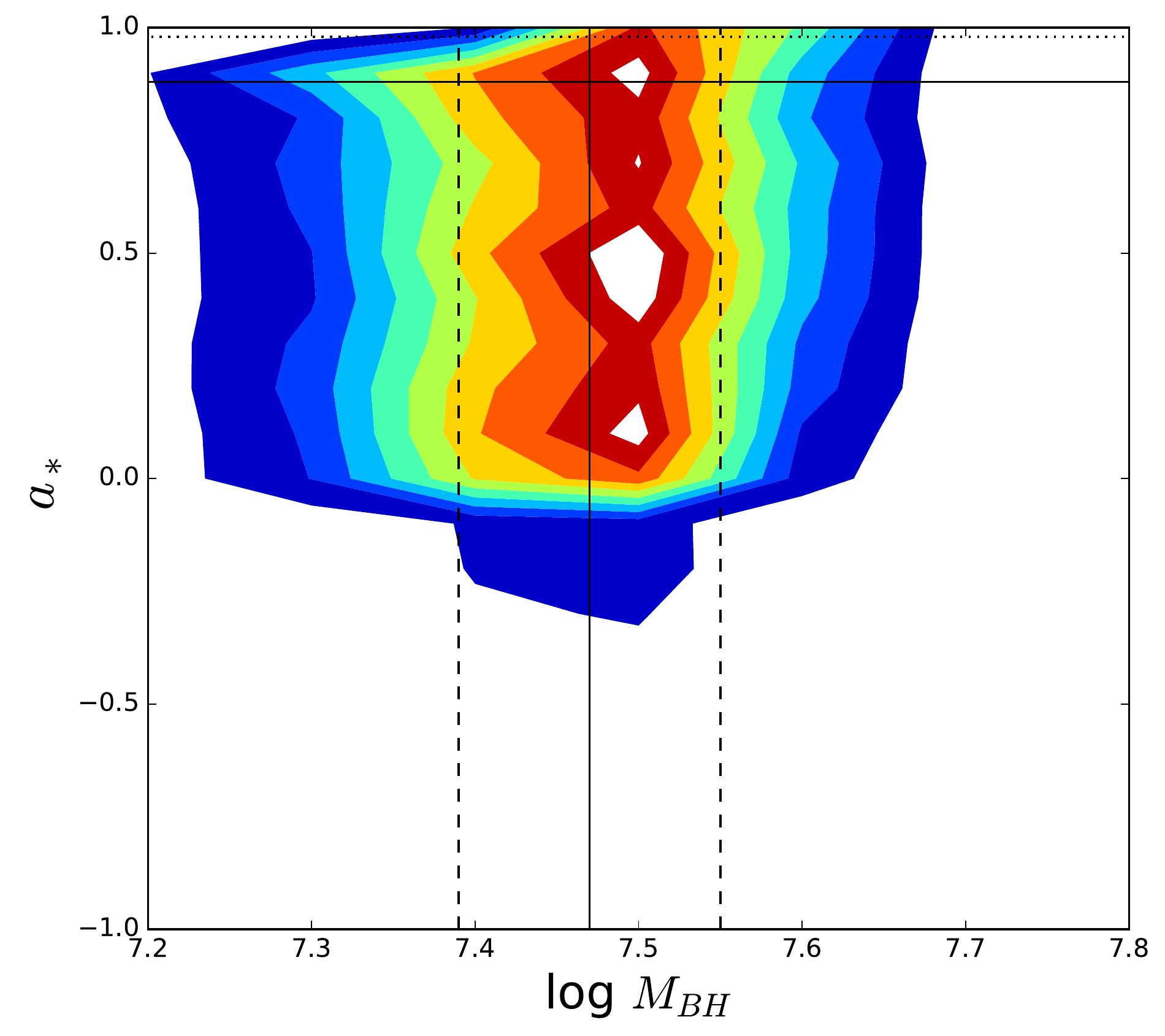}{0.45\textwidth}{}
		  \fig{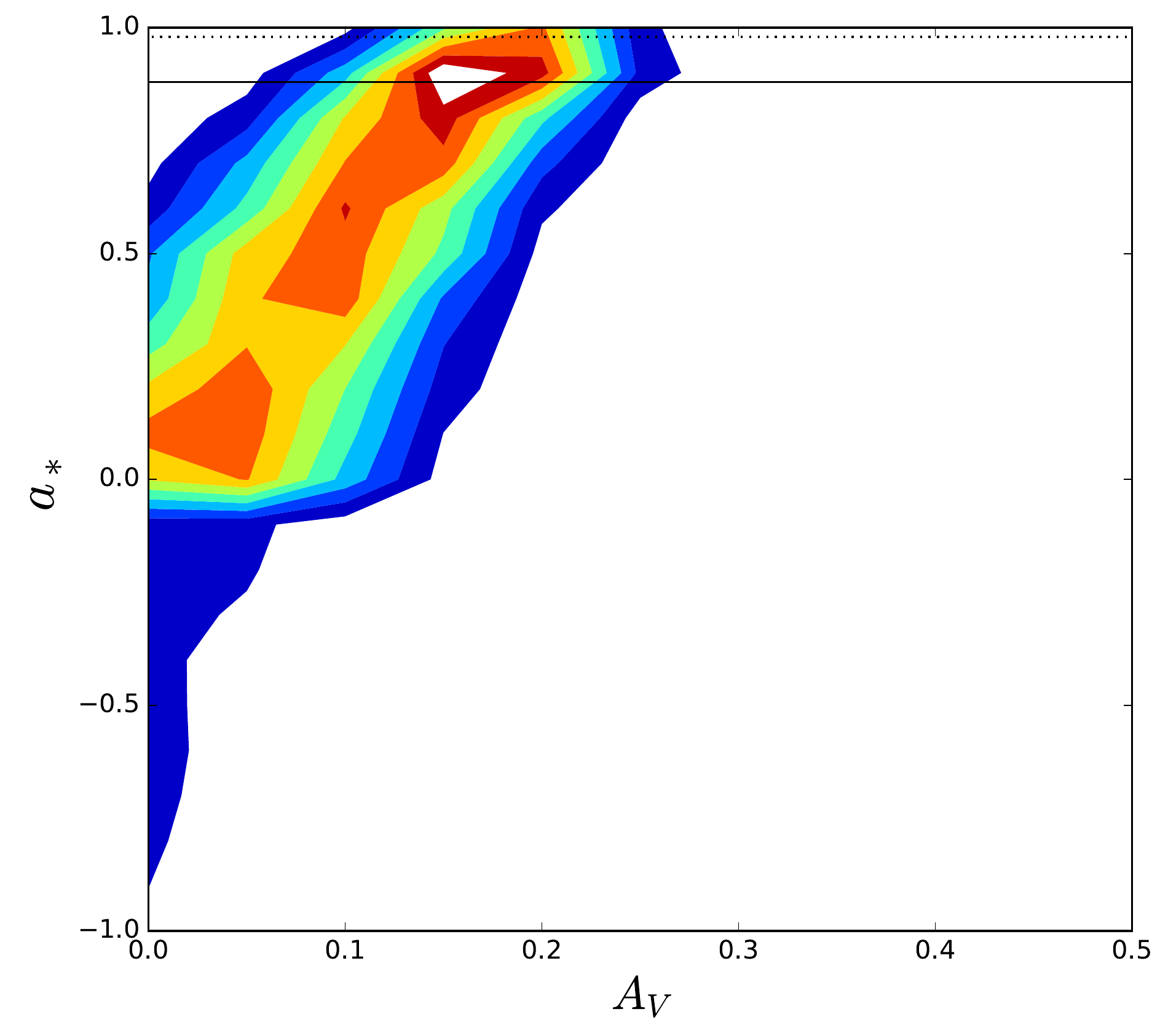}{0.45\textwidth}{}
		  \fig{fig2c.pdf}{0.05\textwidth}{}
          }
\caption{Probability contour plots for NGC 3783 for $a_*$ versus $M_{BH}$ (left
  panels) and $a_*$ versus $A_V$ (right panels) for two different iterations
  of the Bayesian CF routine: without a disc wind (top panels) and with a disc
  wind (bottom panels). The observed value of $M_{BH}$ and associated error is
  indicated by vertical solid and dashed lines, respectively. The horizontal
  dotted and solid lines represent the 90\% and 99\% confidence on $a_*$ from
  B11.
  \label{fig:Ncont}}
\end{figure*}

Even for a maximally spinning black hole, the thin AD emission does not
directly contribute to the hard X-ray band (i.e., above 2 keV; see
Fig.~\ref{fig:Nspec}). Hard X-ray observations of NGC 3783 give a less steep
power law than in the soft X-ray band. For example, T93 find
$\Gamma = 2.14^{+0.24}_{-0.26}$ when applying their warm absorber model to data
from EXOSAT. If we assume that the excess emission indicated by a steeper
powerlaw in the soft X-ray band is due to AD emission, we can subtract the hard
X-ray powerlaw from the soft powerlaw at 0.1 keV to determine a continuum point
for our regular Bayesian CF procedure.

We use a value of 0.1 dex for the error on $M_{BH}$ from \citet{Peterson04}.
The results of the CF routine are presented in the left panel of
Fig.~\ref{fig:Ncont}, and we find a median $a_* \simeq 0.2^{+0.7}_{-0.9}$.
Using the X-ray reflection method, \citet[][hereafter B11]{Brenneman11}
determine a spin parameter $a_* \ge 0.98$ at 90\% confidence and $a_* \ge 0.88$
at 99\% confidence (indicated by horizontal dotted and solid lines in
Fig.~\ref{fig:Ncont}).

\subsection{Applying the CF Method with an AD Wind}

NGC 3783 is known to have a warm absorber in its X-ray spectrum (T93), i.e. an
outflow often presumed to originate from the AD of the AGN
\citep[e.g.,][]{Tombesi13}. If this is the case for NGC 3783, then the thin AD
model must be modified, as the accretion rate would be reduced throughout the
disk as material is ejected.

The \citet{Slone12} thin AD code provides the option of adding a disk wind to
the model. We therefore rerun the CF routine using a model with a self-similar
disk wind, where the mass outflow rate per decade of radius is constant. The
mass outflow rate for NGC 3783 has been estimated to be $\gtrsim160$ times the
accretion rate; however, much of this outflowing gas may come from beyond the
accretion disk \citep[T93;][]{Crenshaw12}. In the absence of an empirical
estimate of the mass outflow rate from the disk itself, we illustrate the
affect of a massive disk wind by choosing a mass accretion rate at the outer
part of the disc equal to three times the accretion rate at the innermost
stable circular orbit (ISCO). The results are presented in the right panel of
Fig.~\ref{fig:Ncont}.

The main difference between these results and the results without the disk wind
is that lower spin parameters ($a_* < 0$) are much less probable in the disk
wind scenario. This arises because the disk wind reduces the accretion rate in
the inner part of the disk and thus suppresses the luminosity at short
wavelengths. Furthermore, while there is a high probability of $a_* \ge 0.88$
both with and without a disk wind, there is clearly a lower probability of
having $a_* \ge 0.98$ if there is no disk wind (there is a factor of $\sim$1.6
difference in radiative efficiency between these two spin parameters). There is
also a positive correlation between the amount of intrinsic reddening and
$a_*$, with $a_* \ge 0.88$ ruled out if there is close to zero reddening.

\section{Discussion} \label{sec:discuss}

Our aim in this work is to compare the derived spin parameters for the X-ray
reflection and CF techniques for two ``case study'' AGN. Table~\ref{tab:table}
summarizes the results of the two methods, including values for $L/L_{Edd}$,
the disk inclination ($\theta$), and instrinsic reddening, as derived from the
CF method.

For H1821+643, a bright AGN with $M_{BH} \sim 2.5 \times 10^9 \Msun$, R14 found
$a_* \gtrsim 0.4$ using the reflection method. For the CF analysis, the HST FOS
spectrum alone is sufficient, and while we do not obtain a very precise
constraint on $a_*$, we find a strong probability of a spin parameter that
exceeds the lower limit from R14, giving consistent results between the
reflection and the CF method. We emphasize here that R14 do not clearly detect
an Fe line, but instead fit excess continuum emission in the soft X-ray. For
some AGN, physical processes other than relativistic reflection are the more
likely cause of this soft excess \citep[][and references therein]{Boissay16},
making this reflection spin measurement a tentative one (see also \S1). For the
CF method, from the posterior probability distribution, it is clear that if
$M_{BH}$ is higher, then $a_*$ would be constrained to the highest allowed
values. Whereas, if $M_{BH}$ is any lower, we would be unable to obtain a
meaningful constraint on $a_*$.

For NGC 3783, the FOS spectrum lies along the power-law portion of the thin AD
model spectrum, and only in the soft X-ray regime can models with different
$a_*$ be distinguished. Fortunately, there is nearly contemporaneous FOS and
ROSAT data for NGC 3783. However, the X-ray data includes the known X-ray
power-law emission that likely originates from above the AD (often called the
``corona''). Using the X-ray flux as an upper-limit, we find that as long as
$M_{BH}$ is at least as high as the observed $M_{BH}$, any spin parameter could
fit the data.

Since there are other components besides the AD emission in the X-ray, if we
assume that just the excess emission indicated by the steeper powerlaw slope in
the soft X-ray, compared to in the hard X-ray, is due to the AD itself,
applying the CF method to NGC 3783 gives a high probability for a high spin
parameter, consistent with the 99\% confidence lower limit from relativistic
reflection (B11). However, there is a low probability of $a_*$ exceeding the
90\% confidence lower limit from B11, unless we include a disk wind in the AD
model.

The results of the CF method are, in general, consistent with the results of
the reflection method for the two AGN studied here. In particular, the
agreement is improved for NGC 3783 if we assume a disc wind, which we include
based on the existence of a warm absorber in the X-ray spectrum. The disk wind
analysis, however, is tentative because it is unknown how much, if any, of the
outflow originates from the inner part of the disk
\citep[see e.g.][]{Netzer03}. If the outflow originates further out and
therefore does not suppress the short wavelength thin AD emission, there is a
slight tension between the two methods, as the reflection method suggests a
slightly higher spin parameter than the CF method without a disk wind. We also
find that, without a disc wind, the highest spins are most probable for $A_V$
between 0.2 and 0.3 mag. While these reddening values are generally consistent
with the constraints from broad emmission line measurements for NGC 3783
\citep[$A_V$=0.1$\pm$0.2;][]{SchnorrM16}, if the reddening is actually closer
to 0.1 mag, then there is even greater tension between the reflection and
CF results for $a_*$. \citet{Done13} and \citet{Done16} similarly find that the
CF method suggests lower spin parameters for narrow-line Seyfert 1s than the
nearly maximal spin typically found for this AGN subclass via X-ray reflection.

Our study highlights one particular strength of the X-ray reflection method for
nearby Seyfert galaxies. For NGC 3783, with a BH mass of $\sim$$10^{7}$ \Msun,
the inability to probe the extreme UV prevents us from obtaining a very precise
estimate of $a_*$. However, we point out that recent work by \citet{Yaqoob16}
casts doubt on whether the Fe K$\alpha$ line gives any information on $a_*$ for
one of the AGN in the reflection sample (see also \S1).

Nearly half (12) of the 25 AGN with spin measurements from the reflection
method have $a_* > 0.9$ \citep{Vasudevan16}. Given that the CF method suggests
lower spin for two cases with near-maximal reflection spin estimates
(1H 0707$-$495 in \citealt{Done16} and NGC 3783 presented here), these high-spin
cases would be good candidates for further comparisons between the reflection
and CF methods, especially those with even lower $M_{BH}$ than NGC 3783, whose
AD SEDs would peak further into the soft X-ray. There is also a new method
proposed by \citet{Chartas16}, based on microlensing, that could be included in
future comparisons of spin estimation techniques.

As more and better X-ray measurements allow the reflection sample to grow and
as better constraints on $M_{BH}$ \citep[see e.g.,][]{Shen15,Mejia16} allow the
CF method to more precisely determine $a_*$ for larger samples of AGN, there
will be a larger population where both methods can be properly applied and
compared. If such comparisons yield good agreement, then each method can be more
confidently applied to the samples they are best suited for -- nearby Seyferts
for the X-ray reflection method and higher redshift quasars for the CF method.
If instead, these comparisons bring further tensions to light, then the
assumptions underlying these methods may need to be revisited.

\acknowledgments

We thank the referee for helpful feedback.
We thank Paulina Lira, Julie Hlavacek-Larrondo, and Helen Russell for useful
discussion.
DMC and DH acknowledge support from a Natural Sciences and Engineering 
Research Council of Canada Discovery Grant and a Fonds de recherche du 
Qu\'{e}bec — Nature et Technologies Nouveaux Chercheurs Grant.
GWF acknowledges support from Universit\'{e} Paris-Saclay's IDEX program and
l'Office Franco-Qu\'{e}b\'{e}cois pour la Jeunesse.



\end{document}
